\documentclass[preprint]{aastex}

\def\lsim{\lower.5ex\hbox{$\; \buildrel < \over \sim \;$}}
\def\gsim{\lower.5ex\hbox{$\; \buildrel > \over \sim \;$}}
\def\lax    {\ifmmode{_<\atop^{\sim}}\else{${_<\atop^{\sim}}$}\fi}
\def\gax    {\ifmmode{_>\atop^{\sim}}\else{${_>\atop^{\sim}}$}\fi}
\def\etal{{\it et al.\/} }
\def\gtorder{\mathrel{\raise.3ex\hbox{$>$}\mkern-14mu
             \lower0.6ex\hbox{$\sim$}}}
\def\ltorder{\mathrel{\raise.3ex\hbox{$<$}\mkern-14mu
             \lower0.6ex\hbox{$\sim$}}}

\def\pmb#1{\setbox0=\hbox{#1}%
  \kern-0.015em\copy0\kern-\wd0
  \kern0.03em\copy0\kern-\wd0
  \kern-0.015em\raise0.0433em\box0 }

\begin{document}

\title{Classification of Power Density Spectrum Features and 
Estimation of the Delta-Invariant Value for the Z Source GX 340+0}

\author{Lev Titarchuk\altaffilmark{1,2}, and 
Vladimir  Osherovich \altaffilmark{3}}

\altaffiltext{1}{George Mason University/CEOSR; US Naval Research Laboratory,
code 7620, Washington, DC. 20375-5352; lev@xeus.nrl.navy.mil}
\altaffiltext{2}{NASA/GSFC, code 661, Greenbelt, MD 20771;  lev@lheapop.gsfc.nasa.gov}
\altaffiltext{3}{NASA/GSFC/Emergent, Greenbelt, MD 20771; vladimir@urap.gsfc.nasa.gov}






\shorttitle{QPOs Classification of GX 340+0}   
\shortauthors{Titarchuk, Osherovich}

\begin{abstract}
We present a theoretical analysis of {\it Rossi X-ray Timing Explorer}  data
of Z source GX 340+0 obtained by Jonker et al. 
In the frameworks of the recently formulated the transition layer
model the $\delta-$angle is an angle between the neutron star (NS) magnetospheric axis and 
the disk (presumably NS rotational) axis.
We determine the angle $\delta=6^o.3\pm 0.^o 3$ which is a combination 
of the simultaneously observed kHz QPO and HBO frequencies.
While these three frequencies change by a factor of three or more
their $\delta-$combination stays almost constant.
   GX 340+0 is the fourth source (in addition to 4U 0614+09, Sco X-1 and  4U 1702-42)
for which  $\delta$ has been determined. 
With at most one (constrained) parameter we make a complete classification of six observed power  spectral
features, including the  two kHz frequencies, the first and second harmonics of 
the HBO frequency, low-frequency noise  component and break frequencies.  We demonstrate that 
a new component discovered  by Jonker et al. in the GX 340+0 power spectrum
is related to  the viscous frequency branch  which has been, in fact 
 reported earlier in 4U 1728-34 by Ford and van der Klis (1998).  
Finally, we re-classify several previously misidentified features in the power spectrum.
\end{abstract}


\section{Introduction}

This {\it Letter} contains  the classification of power  spectral features
in GX 340+0 and a determination of the $\delta-$invariant using
 observational data from Jonker et al. (2000).
Psaltis, Belloni \& van der Klis (1999) discussed many  similarities
and correlations between  the QPO frequencies in atoll sources, 
Z-sources and black hole candidates, which may be related to the similar 
phenomena and physical processes occurring in all of these systems. 
In addition, Wijnands \& van der Klis (1999) found that the break
frequency of the broken power law which describes the power spectrum
correlates well with the frequencies of the peaked noise components observed
in atoll sources and black hole candidates.
Titarchuk \& Osherovich (1999; hereafter TO99)   attempted to 
classify QPO frequencies using the basic properties of  accretion flow 
around  compact objects. The first property is an adjustment of the Keplerian flow
in the disk to the sub-Keplerian inner boundary condition (Titarchuk, Lapidus
\& Muslimov 1998, hereafter TLM98), which is a common phenomenon for neutron star (NS) and black hole 
candidate systems. In this respect
all correlations between the break frequencies $\nu_b$ and low QPO frequencies 
$\nu_{V}$ should be similar in both types of system because they are related to 
the diffusion and oscillation time scales in the transition region between 
the Keplerian flow and the rotational inner boundary 
(either the NS surface or the inner edge of
the accretion flow in BH systems). Later the predicted correlation  was confirmed 
by Ford \& van der Klis (1998) who analyzed RXTE data from 4U 1728-32.
Furthermore, in TO99 the authors demonstrated  that $\nu_b$ and $\nu_{V}$
are related through the power law  $\nu_b\propto \nu_{V}^{1.6}$ which fits to the data 
in most BH and NS systems (Wijnands \& van der Klis, 1999).

The second property is an effect of the sub-Keplerian rotating magnetosphere located
above the disk. Osherovich \& Titarchuk (1999a, hereafter OT99a) suggested
that the radial oscillations of the fluid element bounced  from the disk shock
region (presumably at the adjustment radius) would be seen
as two independent oscillations in the radial and the vertical directions
due to the presence  of Coriolis force in the magnetospheric rotational frame 
of reference. The observed evolution of two kHz peaks allows one to infer
the profile of the differentially rotated magnetosphere and then to verify
it using the observed correlation of the horizontal branch quasi-periodic oscillations 
(HBO) frequencies  with the low kHz peak.

In a black hole system the corona above the disk should  also rotate  with  less
than Keplerian velocity.  When oscillating fluid elements of the disk are thrown to the
corona,  these oscillators would be under the influence of the  Coriolis force which leads
to the same rotational splitting effect (OT99a) seen in the NS systems.
The main QPO Keplerian  frequency would be split  into two frequencies, the hybrid and low branch
related to the radial and vertical oscillations respectively in the rotational frame of references. 

The recent discovery of a 450 Hz QPO (Strohmayer 2001) in addition to
the previously reported 18 and 300 Hz (Remmilard 1999) in GRO J1655-40
provides additional support for the rotational splitting effect.

There are three other QPO models in the literature. The first one is the beat frequency model 
suggested by Alpar \& Shaham (1985), and recently modified as a sonic-point beat frequency 
model by Miller, Lamb \& Psaltis (1998). The second  is the relativistic precession
model proposed by Stella \& Vietri (1998, 1999). The third  is the inner accretion disk
model promoted by Psaltis \& Norman (2000). All observational arguments for these models
and their problems are discussed in detail  by Wu (2001) where these models are also
compared with the transition layer model (TLM) (TLM98, TO99-OT99a).
We do not go into the details of this comparison in this Letter but we do give more arguments
for TLM in terms of the new QPO data for GX 340+0 (Jonker et al. 2000, here after J00).  
     
Precise simultaneous measurements of the frequencies of the two
kHz QPOs and HBO harmonics in a wide frequency range for GX 340+0 (J00) 
give us an opportunity to derive the $\delta-$invariant.
In Osherovich \& Titarchuk (1999, hereafter OT99b), 
the authors demonstrated for the source 4U 1702-42 
that the inferred angle $\delta$, (see Eq 2, 5 in OT99b) 
\begin{equation}
\delta=\arcsin\left[(\nu_h^2-\nu_{\rm K}^2)^{-1/2}
(\nu_L\nu_h/\nu_{\rm K})\right]
\end{equation} 
 depends only  on  
{\it observed frequencies:}  the low and high kHz QPO peaks  
$\nu_{\rm K}$, $\nu_h$ and  the HBO (the low branch) peak $\nu_L$. It stays the same  
  ($3.9^o\pm0.2^o$) over significant range of $\nu_{\rm K}$ (650-900 Hz). 
Using  a series of observations where $\nu_h$, $\nu_{\rm K}$, $\nu_L$
 are  detected simultaneously [van der Klis et al. 1997; 
Markwardt, Strohmayer \& Swank 1999; van Straaten et al. 2000; J00)]  we check that  the angle $\delta$ is a true invariant  
for  four specific sources: Sco X-1, 4U 1702-42, 4U 0614+09 and GX 340+0 (see \S 2). 
 We  present the magnetospheric rotational profile for GX 340+0  
 inferred from  J00 data in \S 3.  Using this profile  and the $\delta$ value 
we will reproduce the theoretical dependence of the low branch on the low kHz
frequency in \S 4 and compare it with the J00 observations.   

According to the TLM, also known
as ``two-oscillator model'' all frequencies (namely $\nu_h$, $\nu_L$, $\nu_b$
and $\nu_{V}$) have specific dependences on $\nu_{\rm K}$.

The comprehensive summary of 
the TML model and its successes to date is present in Titarchuk \& Osherovich (2000a)
 
In \S 5  we offer  the full classification of the QPO frequencies
containing six branches which fit to the data with at most one (constrained) 
parameter.  Discussion and summary follow in the last section.

\section{$\delta-$Invariant and Verification of Transition Layer Model}
We have used the frequencies measured for sources  GX 340+0, 4U 0614+09, Sco X-1 
and 4U 1702-42 where the low and high kHz QPO peaks $\nu_{\rm K}$
and $\nu_h$  and HBO frequencies $\nu_L$   are  measured simultaneously.
The resulting values of $\delta$ calculated from Eq. (1) are shown in Figure 1. 
Indeed, for  each of these sources, the  $\delta-$values
 show little variation with $\nu_{\rm K}$, $\nu_h$, $\nu_L$.
They are $3^o.9\pm 0^o.2$,  $5^o.5\pm 0^o.5$, $15^o.3\pm 0^o.5$, for 4U 1702-42, 
Sco X-1 and 4U 0614+09 respectively.
The angle $\delta$ obtained  for the source  GX 340+0 is 
$6^o.3\pm0^o.3$ and is similar to the values obtained for Sco X-1 and 4U 1702-42.
The existence of   the $\delta-$invariant  predicted by the TLM 
[OT99b, Titarchuk \& Osherovich (2000b here after TO00)] is a 
challenge for any other QPO model. It is important to note that   
{\it all frequencies included in this $\delta-$relation are observed 
frequencies and thus the relation Eq. (1) is a model independent invariant}.
It is worth pointing out once again that the $delta-$angle varies from source to source.

In fact, Wu (2001) confirms the existence of the delta-invariant for three more sources -- Cyg X-2, 
GX 17+2 and GX 5-1, using the RXTE observational data obtained by Psaltis et al. (1999),
Wijnands et al. (1998a,b, 1997). For all these sources the $\delta-$angle is about
$6^o$ which is very close to that obtained for Sco X-1 and GX 340+0. 
The accuracy of $\delta-$ determination is of order  5\%.

\bigskip
\par
\noindent
\section{ Inferred Rotational Frequency Profile of the NS Magnetosphere in
GX  340+0}

From the observed kHz frequencies (J00, Table 3), $\nu_h$
and $\nu_{\rm K}$, the profile of $\nu_{mag}=\Omega(\nu_K)/2\pi$ has been calculated 
according to 
\begin{equation}
 \nu_h=[\nu_{\rm K}^2+4(\Omega/2\pi)^2]^{1/2},
\end{equation}
 and modeled using  the theoretically  inferred magnetic multipole structure of 
 a differentially rotating magnetosphere (OT99a) 
\begin{equation}
\nu_{mag}=C_0+C_1\nu_{\rm K}^{4/3}+C_2\nu_{\rm K}^{8/3}+C_3\nu_{\rm K}^4
\end{equation}
where $C_2=2(C_1C_3)^{1/2}$. The constants
$C_0\equiv\nu^{0}_{mag}=330$ Hz, $C_1=-9.72\times 10^{-2}$ Hz$^{-1/3}$,
$C_2=5.34\times10^{-5}$ Hz$^{-5/3}$ and $C_3=-7.32\times10^{-9}$
Hz$^{-3}$ have been obtained  by a least-squares fit with $\chi^2=15.8/
12$. 
The $\nu_{mag}$ profile  for GX 340+0 is  
very similar to those of 4U 1608-52, and Sco X-1 (see OT99a, Figs. 1-2).

\section{ Low Branch Frequency vs kHz QPO Frequency}
The GX 340+0 data of J00 allows one to  check the prediction of the model
for the low branch frequency $\nu_{L}$. Taking the profile of $\nu_{mag}(\nu_{\rm K})$, 
(Eq. 3), we plot $\nu_{L}$ in Figure 2. Fixing $\delta=6^o.3$, we find
that  our plot for $\nu_L$ and $2\nu_L$, calculated according to formula (OT99a, Eq. 9) 
\begin{equation}
\nu_L=2\nu_{mag}(\nu_{\rm K}/\nu_h)\sin\delta,
\end{equation} 
 fits the data for the observed frequencies
of (20-40) Hz and (40-80) Hz respectively. 
There are three observational points in Figure 2 (square points with error bars) which
we interpret as the second harmonics $2\nu_L$. The first harmonics $\nu_L$ for these frequencies
are not observed by J00 and they are obtained using $2\nu_L$ and presented as squares without 
error bars.
The error bars of the observational points shown by circle marks  are less than the mark size in
most cases.   
It is worth noting that the observational points follow the theoretically predicted curve, 
which shows the signs of saturation at high $\nu_{\rm K}$. The same type of behavior 
of $\nu_L$ vs $\nu_{\rm K}$ is also seen in the Sco X-1 data.   

\section{ Break and Viscous frequencies vs kHz QPO Frequency Correlations}

Further tests of the TLM  can be done using a 
comparison of the observed correlations of break and viscous  QPO frequencies
vs  kHz QPO frequencies (J00) with the theoretical dependences derived in TLM98 and TO99.
In Figure 3  (the classification plot),  we present the theoretical curves calculated 
using Eqs. (9, 12-13) in TO99 using the dimensionless parameter 
$a_{\rm K}=m(x_0/3)^{3/2}(\nu_0/363~{\rm Hz})$ where  $\nu_0$ is the neutron star (NS) spin, 
$m$ is the NS mass in the solar masses, $x_0$ is the NS radius in
units of Schwarzchild radii.
In TO00 (Eq. 4) we approximate the solution of Eq (9) of 
TO99 by a polynomial $\nu_b=C_b P_4(\nu_{\rm K})$ for $a_{\rm K}=1.03$ and 
$\nu_{0,363}= \nu_0/363~{\rm Hz}=1$.
The normalization of the theoretical curve is controlled by the constant $C_b$ (TO00) which 
reflects the properties of the specific source.
 For instance, for 4U 1728-34  the observed frequencies are  fitted by the curve 
(TO99) for which the constant $C_b=1$ 
and $R_0=10$ km for 1.4 solar masses (see TO00, Eq. 2). It should be pointed out that the NS spin  for
4U 1728-34 is known to be $\nu_0= 363$ Hz.    
For GX 340+0 we fit the data points  by  curves with 
the parameter $\nu_0=363$ Hz  for which  the corresponding constant $C_b=7.6$. 
For the source GX 340+0 we found the Lebesgue's measure 
(for a definition, see Titarchuk, Osherovich \& Kuznetsov 1999, hereafter TOK) is 16\% for $a_{\rm K}=1.03$. 
In fact, the  theoretical curves are functions of two parameters,
$a_{\rm K}$ and $\nu_0$, and for (a given) $\nu_{0,363}$ that curve with $\nu_{0,363}=1$ can be 
obtained by scaling the argument (see TO00).    

According to solutions presented in TO99 and TOK, the break frequency $\nu_b$  
correlates with the low Lorentzian frequency $\nu_{V}$ (which we 
interpreted as a viscous frequency in the TL model) through the power law relation 
\begin{equation}
\nu_b \propto\nu_{V}^{1.61}.
\end{equation}
 Using the data of Wijnands \& van der Klis (1999) it was confirmed in TOK that approximately  
the same index   is valid for most NS and 
BH objects.  We have also found that the observed correlation 
of $\nu_b$ vs $\nu_{V}$ in GX 340+0 is well described  by the same 
power law dependence (Eq. 5) but with a proportionality coefficient $0.08$ (which is twice more than 
that for 4U 1728-34).
The dependence of $\nu_V$ vs $\nu_{\rm K}$ is shown in Figure 3 by the red curve. 
To fit  the theoretical curve to the data we use only those  data points 
which were specified by J00 as
the new discovered frequencies. They pointed out that these frequencies well close to half that of
the HBO.  
However looking at the classification in Figure 3 we argue that these power spectral features 
are the same low noise component frequencies -- in our terms, $\nu_V$-- that were found 
by Ford and van der Klis (1998) in 4U 1728-34 and by TOK in Sco X-1. Furthermore, we have found that at least 
two power spectra features which were identified
by J00 as the break frequencies also belong to the viscous branch. The appropriate points are denoted
by triangle marks in Fig. 3.  

The  dependences of $\nu_b$ and $\nu_V$ on $\nu_{\rm K}$ 
allow us to retrieve information regarding the turbulent scale $l_{\rm fp}$, 
the sonic velocity $v_s$, the magnetic field strength and structure in the transition layer. 
We discuss these issues in detail in Titarchuk, Bradshaw \& Wood (2001).

Wu (2001) uses the $\alpha-$prescription -- i.e. $\alpha\sim (l_{\rm fp}v_t)/(v_sH)$, where $H$ is
a half-thickness of the disk -- 
to specify the absolute value of the viscous frequency
$\nu_V$. However we   argue that the introduction of one more free parameter
($\alpha$) 
in TL model is not necessary and furthermore it leads to  very small values of  
 kinematic viscosity, $\mu_{\rm kin}=v_tl_{\rm fp}/3$. 
One may  conclude  using Wu's 
best-fit parameters of $\alpha(H/R)^2\sim 10^{-3}$ that $v_tl_{\rm fp}$  is at least two
orders of magnitude less than $ v_rL$.
On the other hand, the Reynold's number $\gamma$ is tightly constrained
by the observed absolute values of kHz frequencies (TLM98, Fig. 3) and one can find that
$ v_tl_{\rm fp}> Lv_r/4$.  In TLM98 authors have already realized this problem with the $\alpha-$
prescription and therefore they have not used it in the model.

\section{Discussion and Conclusions}

In this {\it Letter} we have further  verified predictions of the transition layer (TL) 
model for  four sources by presenting our analysis of Jonker et al.'s GX 340+0 data. 
Furthermore we demonstrate that the right classification of the power spectra features
allows us to shed light  on their nature and origin. 
The unifying characteristic of TLM is that all QPO features are natural and  necessary
consequence of the adjustment of Keplerian disk to the sub-Keplerian rotation 
of the neutron star. In the transition layer the interaction of the two oscillators may occur 
because they share  the common boundary at the outer edge of the  transition 
layer. In this case  a strong dependence  is supposed to exist between all frequencies in the TLM 
and the fundamental Keplerian frequency $\nu_{\rm K}$ at the TL outer edge. 
In our model, assuming 
high electric conductivity, we describe the frequency of the differential 
rotation of the magnetosphere $\nu_{mag}$. 
We also assume that the 
NS magnetic field has a discontinuity 
in the equatorial plane similar to the  discontinuity of the field in the 
equatorial plane of the solar corona and related current sheet in the 
heliosphere as described in the model of Osherovich, Tzur and Gliner (1984).
Both the current sheet and differential rotation are observed elements 
of solar corona and heliosphere (e.g. Hoeksema \& Scherrer 1987, Burlaga 1995). 

The TLM model  reveals  the physical nature of the low branch (HBO)
oscillation frequency as a low branch  of the Keplerian oscillator 
under the influence of the Coriolis force. 
The angle $\delta$ as a global parameter describes the
inclination of the magnetospheric equator to the equatorial plane of the 
disk. Measured locally for different radial distances
(therefore different $\nu_K$), $\delta$ may vary considerably if 
the observed oscillations do not correspond to the predicted low
Keplerian branch. The angle $\delta$ is calculated using the observed frequencies only.  
The constancy of $\delta$ shown in Figure 1 argues in favor of our 
model.   Thus, if a similar rotational splitting effect occurs
in BH systems, are the needed frequency trio simultaneously
observed: a pair of high frequencies of order 200-300 Hz (for
 black hole with mass of 7-10), and a low frequency close to 20 Hz?
Strohmayer (2001) reports precisely this case for BH candidate
GRO J1655-40.  He simultaneously observed three frequencies, 450, 300
and 18 Hz.

If we assume the rotational splitting paradigm as
an origin for these three frequencies, we can estimate the rotational
frequency of the corona-configuration above the disk $\Omega/2\pi=
167.7 $ Hz using the hybrid relation (OT99, Eq. 7) for  $\nu_h=450$ Hz
and $\nu_{\rm K}=300$ Hz. Furthermore, if  we assume
 that the corona configuration 
rigidly rotates with  
the black hole, we can estimate the BH dimensionless angular momentum $a=cJ/GM^2=\Omega
GM/c^3=3.6\times10^{-2}$. We have also assumed
a BH mass of about seven solar masses.  The low 18 Hz frequency can be
interpreted as the low branch QPO if $\delta=4.6^o$ (see Eq. 1), which is a typical 
for NS systems (see Fig. 1).

 Strohmayer (2001) argues that the GRO J1655-40 QPO data, with
$0.4<a<0.6$, could account for the higher frequency, 450 Hz, when identified with
the radial epicyclic frequency within relativistic disk models (see e.g.
 Stella, Vietri \& Morsink 1999; Psaltis \& Norman 2000; Markovich 2000).
 He also suggests that the 300 Hz QPO is associated with lower kHz
QPOs in neutron stars ({\it c.f.} OT99 and Psaltis et al 1999).  However, if this parallel
between NS and BH systems assumes that the higher frequency QPO twin frequency is due to
General relativity (GR) effects, the question arises of how this GR model
 works for NS systems when kHz QPO frequencies of 400-500 H (corresponding
 to 10-12 Schwarzschild radii where GR effects are negligible) are observed.
 Such QPOs, along with higher kHz frequencies, are seen very often in
NS systems (see Fig. 4). One may conclude that  the association between QPOs
in NS and BHC along with their interpretation within the GR model is somewhat
problematic.

The observed correlations of the viscous and break frequencies versus kHz
QPO frequencies allow us to retrieve information about the NS spin. We found that for GX 340+0 as
well as for Sco X-1 the NS spin  should be close to 360 Hz which is  similar to that measured directly 
in 4U 1728-34 (Strohmayer et al. 1996). On the other hand there is an indication (TO00) that
the NS spin in 4U 0614+09 can be  as high as 700 Hz  or more.

The authors acknowledge discussions with
Joe Fainberg, Kent Wood and Reba Bandyopadhyay.  We are grateful to  
Peter Jonker  for use of his data. We are also grateful to the referee for a thorough analysis
of the paper content and constructive suggestions

\clearpage

\begin{figure}
\caption{ $\delta-$invariant. 
Inferred angles $\delta$ between the rotational axis of 
magnetosphere and the normal to the plane of Keplerian oscillations as a 
function of low peak kHz QPO frequency for four sources: GX 340+0 (green), 4U 0614+09 (magenta), 
Sco X-1 (blue), and  4U 1702-42 (red) .  The $\delta-$invariant (eq.[1]) is calculated using the 
observational frequencies $\nu_h$, $\nu_{\rm K}$ and $\nu_L$ only.
\label{Fig.1}}
\end{figure}


\begin{figure}
\caption{ HBO frequency  vs low kHz QPO frequency for GX 340+0 (Jonker et al. 2000).  
The solid  line is the theoretical dependence of the low branch oscillation  frequencies (black)
 and  the second harmonics (blue) on the Keplerian  frequency, calculated using multipole expansion 
 of the rotational frequency. The points   for the first harmonic of the 
 low branch (green squares) are obtained by dividing by 2 the observational frequencies of 
 the second harmonic  (which are the green squares with error bars). 
\label{Fig.2}}
\end{figure}

\begin{figure}
\caption{Classification of QPOs in the Z source GX 340+0.
 Solid lines are the theoretical curves: Blue for the Upper hybrid frequency 
branch, magenta for the second harmonic of the low branch, blue  for the first harmonic
of the low branch, red  for the viscous branch and black  for the break frequency branch.
The points with squares for the first harmonic of the low branch are obtained by dividing by 2 
the observational frequencies of the second harmonic (blue squares).    
The observational points with black triangles are identified by Jonker et al. (2000) 
as break frequencies.
The theoretical curves for viscous  and break frequencies branches are constructed using 
the first six viscous frequency points (red asterisks) and the first seven break frequency
 points (black triangles) respectively. 
\label{Fig.3}}
\end{figure}

\noindent
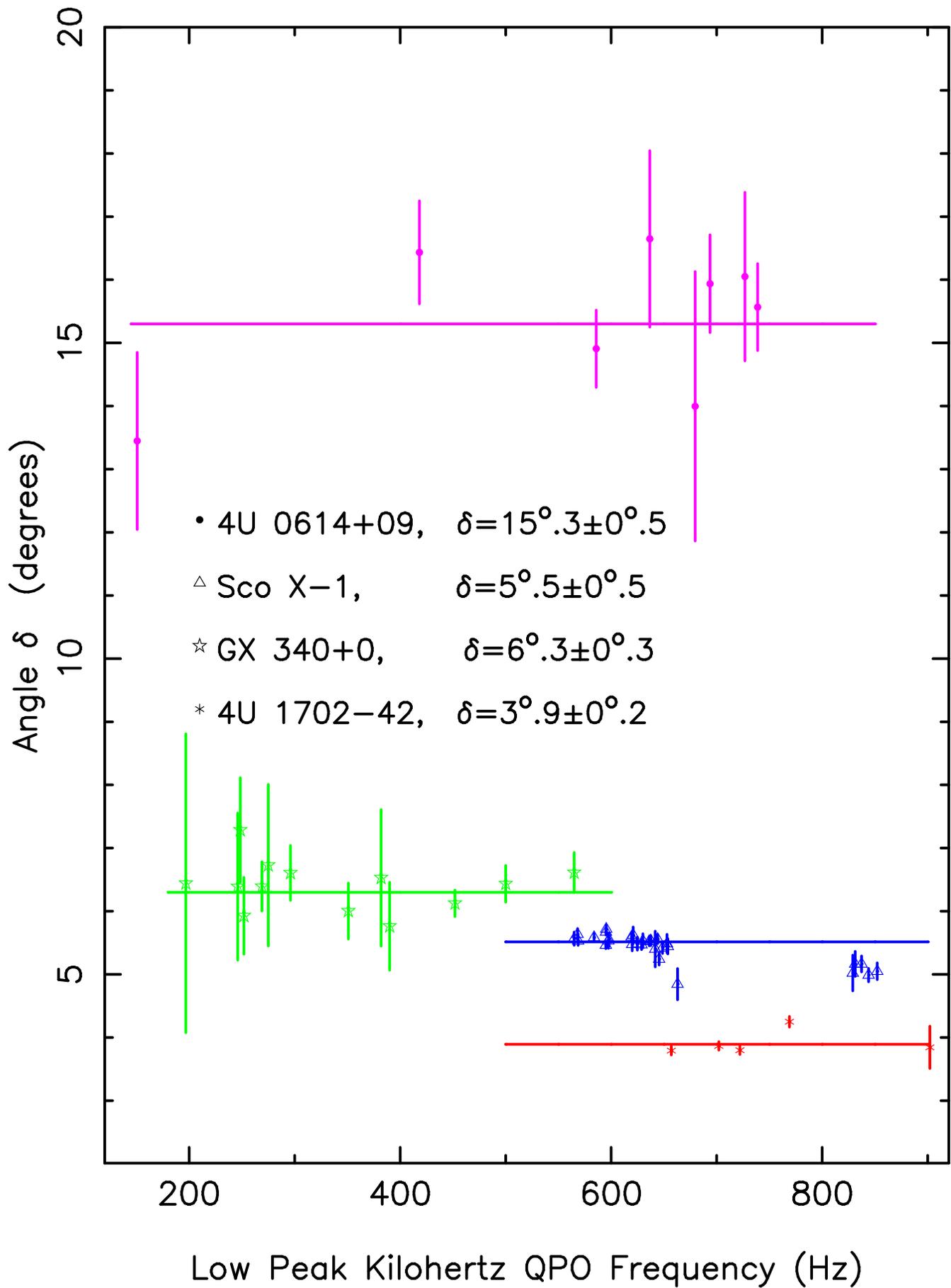
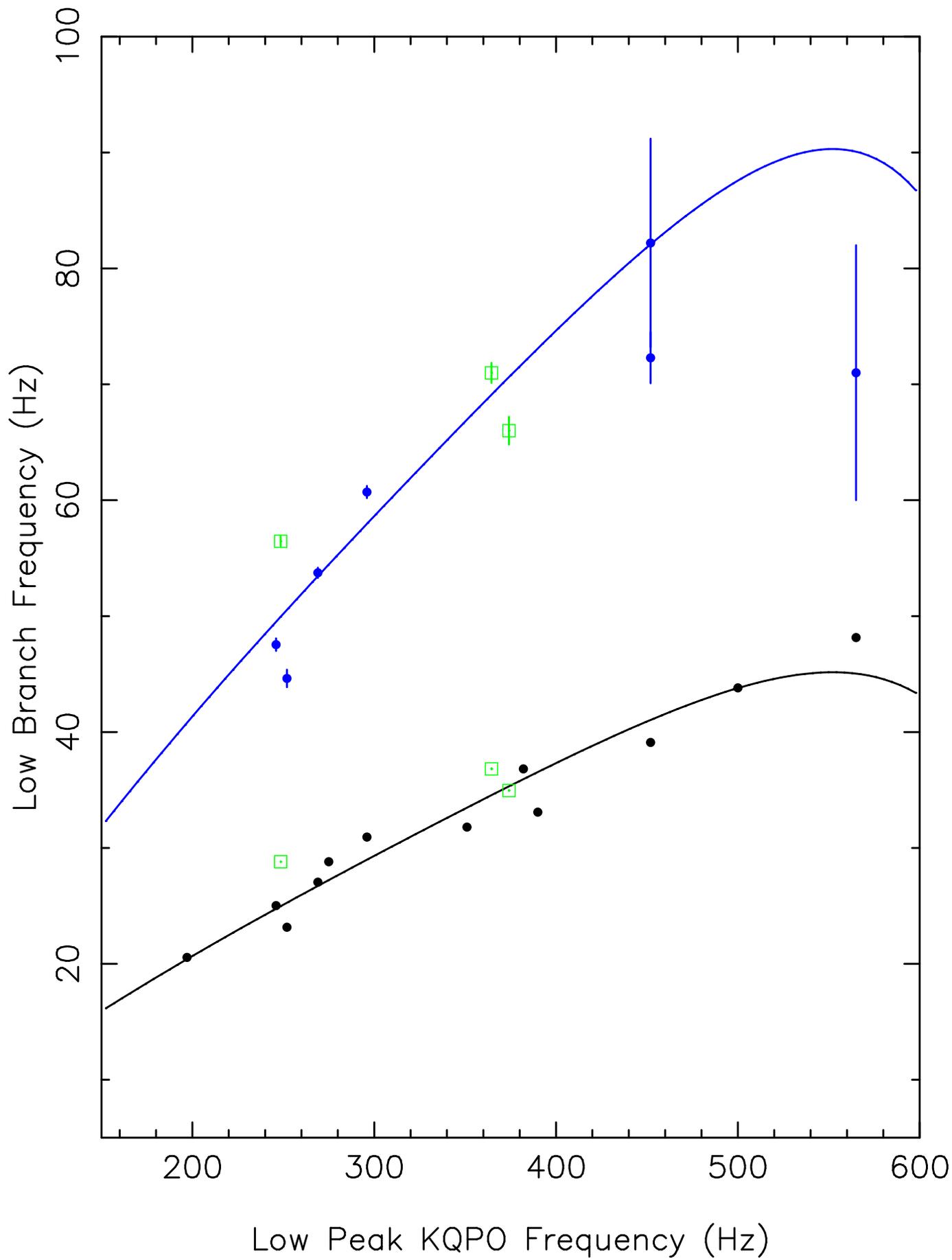
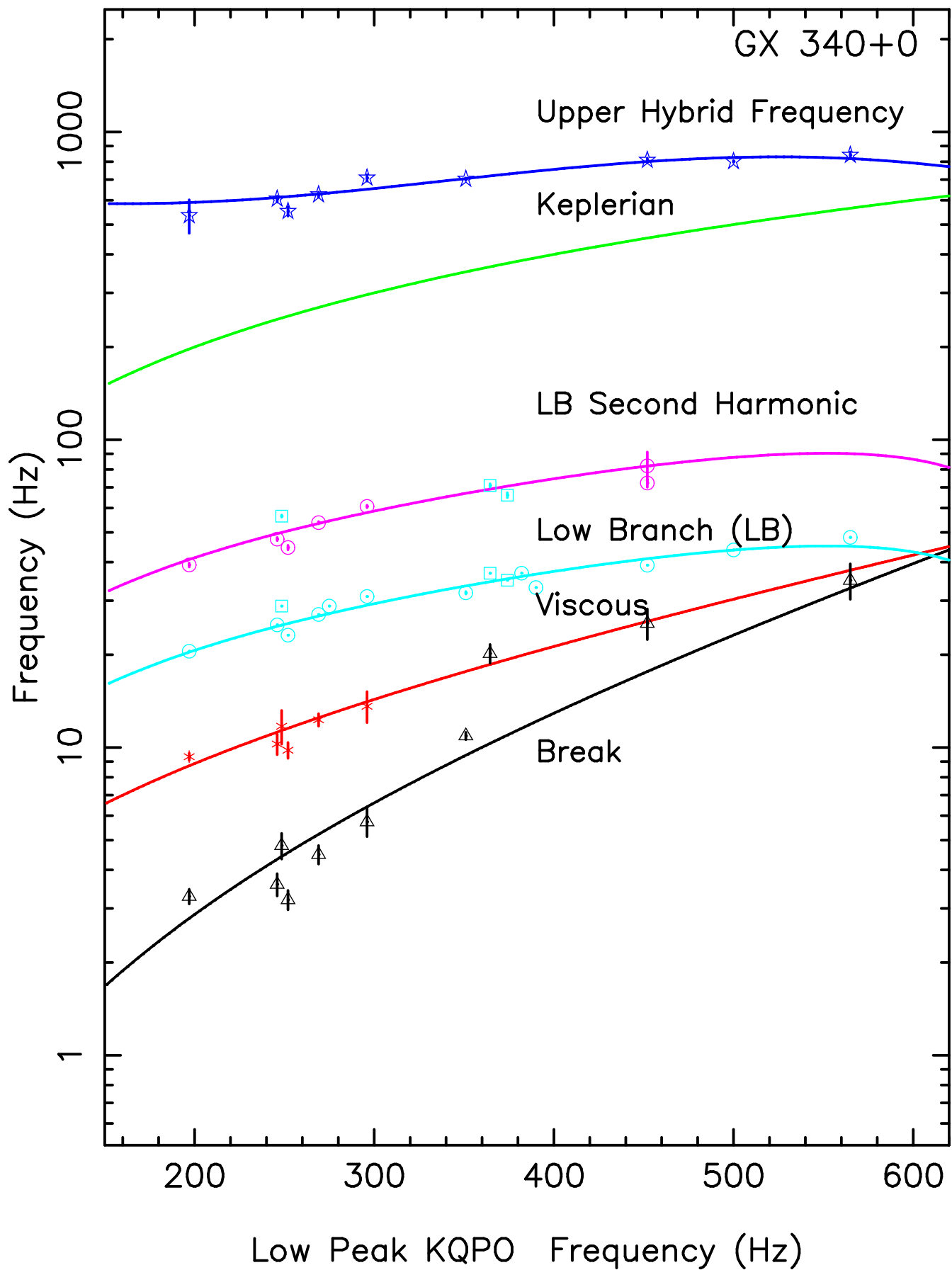
\end{document}